\documentclass[10pt, conference, compsocconf]{IEEEtran}
\usepackage{amsmath, amssymb}
\usepackage{graphicx}
\usepackage{footnote}

\begin{document}
\title{Application Delay Modelling for Variable Length Packets in Single Cell IEEE 802.11 WLANs}

\author{
\IEEEauthorblockN{Albert Sunny\IEEEauthorrefmark{1}, Joy Kuri\IEEEauthorrefmark{2} and Saurabh Aggarwal\IEEEauthorrefmark{3}}
\IEEEauthorblockA{Center for Electronics Design and Technology\\
Indian Institute of Science, Bangalore-560012, India\\
Email: \IEEEauthorrefmark{1}salbert@cedt.iisc.ernet.in, \IEEEauthorrefmark{2}kuri@cedt.iisc.ernet.in, \IEEEauthorrefmark{3}saggarwal@cedt.iisc.ernet.in}
}

\bibliographystyle{IEEEtran}

\maketitle

\begin{abstract}
In this paper, we consider the problem of modelling the average delay experienced by an application packets of variable length in a single cell IEEE 802.11 DCF wireless local area network. The packet arrival process at each node $i$ is assumed to be a \emph{stationary and independent increment} random process with mean $a_i$ and second moment $a^{(2)}_i$. The packet lengths at node $i$ are assumed to be i.i.d random variables $P_i$ with finite mean and second moment. A closed form expression has been derived for the same. We assume the input arrival process across queues to be uncorrelated Poison processes. As the nodes share a single channel, they have to contend with one another for a successful transmission. The mean delay for a packet has been approximated by modelling the system as a 1-limited Random Polling system with zero switchover times. Extensive simulations are conducted to verify the analytical results.	
\end{abstract}


\IEEEpeerreviewmaketitle

\section{\large{Introduction and Related Work}}
\label{intro}

The IEEE 802.11 has become ubiquitous and gained widespread popularity as a protocol for wireless networks. As a result, various models have been proposed to analyze and model the parameters of interest.

Since the seminal paper by Bianchi \cite{bianchi}, throughput analysis of IEEE 802.11 DCF has come under much scrutiny. In \cite{bianchi}, the main feature of the analysis is the 2-dimensional Markov model, which captures the back-off phenomenon of IEEE 802.11, given a transmission attempt rate for each node. In \cite{tay}, the authors give an analytical model for throughput analysis of DCF using average back-off state as compared to the Markovian model being proposed by Bianchi. In \cite{akumar}, the authors study the fixed point solution and performance measure in a more generalized framework.

Delay analysis of IEEE 802.11 DCF is limited in comparison to the throughput studies. In \cite{tobagi}, the authors propose System Centric and User Centric Queuing Models for IEEE 802.11 based Wireless LANs. In the System Centric Model, the arrivals are assumed to be Poisson, thus the resource sharing model takes the form of an M/G/1/PS system with the mean delay being the same as that in an equivalent M/M/1 system. In the User Centric  Model, each user queue is modeled as a separate G/G/1 queue. 

In \cite{panda}, the authors provide an analysis of the coupled queue process by studying a lower dimensional process and by introducing a certain conditional independence approximation. The authors in \cite{panda}, provide an analytical framework to model the delay only for the case of homogeneous Poisson arrivals. In \cite{joy}, we have analyzed the mean delay for single hop wireless mesh networks under light aggregate traffic by introducing a decoupling assumption. But as load increases, interactions between the queues appear and our modelling assumption ceases to be valid. In \cite{tickoo}, the authors analyze delay under homogeneous arrivals assuming packet lengths to be i.i.d across all the queues. In our previous work \cite{saurabh}, we have addressed delay modelling for nonhomogeneous Poisson arrivals under the assumption of fixed packet length.

In this paper, we model the system as a 1-limited random polling system with zero switchover times. We provide a simple model to obtain the mean delay for the variable length packet arrival process at node $i$. This enables us to use the mean delay expressions from \cite{lee} to analyze the delay in a single cell wireless local area network. We remark that the user traffic delay is not merely the Head-Of-Line (HOL) packet delay that has been analyzed in \cite{bianchi}, \cite{tay} and \cite{akumar}; it includes the delay from the time a user packet arrives at the queue, till the packet reaches the destination. Thus, both queuing delay and HOL delay are included. 

Our objective is to explore the use of known results for the \emph{saturated} network and \emph{random polling systems} in analyzing the mean delay experienced by a packet. Our main contributions can be outlined as follows,
\begin{itemize}
\item We propose a random polling system framework to analyze mean delay in a single cell wireless local area network with variable packet lengths.
\item We obtain closed form expression using a novel approach for mean delay by applying results from \cite{lee} to our random polling system framework, for the case when the packet lengths at node $i$ are i.i.d random variables.
\item We show though simulations that our random polling framework can be used to estimate the mean delay in a single cell IEEE 802.11 wireless local area network.
\end{itemize}

The rest of the paper is organized as follows. In Section II, the system model is described in detail. In Section III, we discuss the delay modelling framework in detail. In section IV,  we obtain closed form expressions for application level mean delay under Poisson packet arrivals and random packet lengths. In Section IV, the proposed framework is validated against simulation results. Finally, Section V presents conclusion and remarks regarding future work.

\section{\large{System model}}
\label{model}
\begin{figure}[h]
\centering
\includegraphics[scale=0.3]{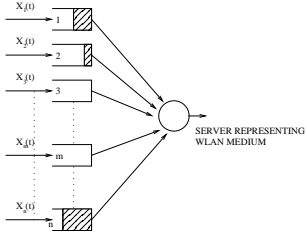}
\caption {System model}
\label{fig:model}
\end{figure}

We consider a single cell wireless local area network with $n$ nodes and no hidden nodes. We assume that the packet arrival process at node $i$ to be a \emph{stationary and independent increment} random process $X_i(t)$ with mean $a_i$ and second moment $a^{(2)}_i$. The lengths of packets at node $i$, are assumed to be i.i.d random variables represented by $P_i$ having finite mean and second moment. Each node in the network shares the medium and uses the IEEE 802.11 DCF to exchange data with one another. A packet transmitted by a node is destined for any of the other nodes. Each node is assumed to have a single network output queue. All wireless links are assumed to operate at the same date rate.
We assume the network layer to employ fragmentation to break down application packets that have length greater than the MTU, since the maximum size of data that can be transmitted in 802.11 at a time is bounded above by the size of the MTU.

From \cite{bianchi}, \cite{tay} and \cite{akumar}, it is known that the average aggregate rate of data transfer is dependent on the number of nodes contending. We also note that each node has equal probability of success. As in \cite{tobagi}, we model the network as a multiple queue single shared server system, where the service rate of the server is dependent on the number of non-empty queues and the server selects a non-empty queue uniformly at random. We assume that the destination can push data out of the network instantaneously. In this system, we are interested in quantifying the average delay between a packet's arrival to a queue and its departure from the system. 

\section{\large{Analysis Framework}}
\label{analysis}

\subsection{Computing aggregate service rate in 802.11 DCF}
\label{bia}
In our model, atmost $n$ nodes contend for access to the wireless medium. Let $\beta_n$ be the probability that a node attempts transmission, when $n$ nodes are contending for access to the wireless medium. From \cite{bianchi}, $\beta_n$ can be expressed in terms of the conditional collision probability $p$ in two ways as follows.

\begin{eqnarray}
\beta_n(p) &=& \frac{2 \cdot (1-2p)}{(W+1) \cdot (1-2p) + pW \cdot (1 - {(2p)}^m)} \label{eq:beta1} \\
\beta_n(p) &=& 1 - (1-p)^{\frac{1}{n-1}} \label{eq:beta2}
\end{eqnarray}

Now, the RHS of Equation \eqref{eq:beta1} is monotonically decreasing from $\frac{2}{W+1}$ to $\frac{2}{2^mW+1}$, for $p \in [0,1]$, and the RHS of Equation \eqref{eq:beta2} is monotonically increasing from $0$ to $1$, for $p \in [0,1]$. We can use fixed point analysis to obtain $\beta$ for a given $n$. Similar approach has been followed in \cite{akumar}. Let $T_S, T_I$ and $T_C$ be the durations of success, idle and collision slots, respectively. Now, $T_S$ can be represented as $\frac{(H + E[P])}{R}$, where $H$ and $E[P]$ denote the header and expected packet lengths respectively with $R$ being the rate of transmission. Let us define the probability of a successful transmission ($p^{(n)}_S$), collision ($p^{(n)}_C$) and idle ($p^{(n)}_I$) as
$$ p^{(n)}_S = n\beta_n \cdot (1-\beta_n)^{n-1} \quad ;\quad p^{(n)}_I = (1-\beta_n)^n $$
$$ p^{(n)}_C = 1-  p^{(n)}_S - p^{(n)}_I$$	
By applying the Renewal Reward theorem, we define the aggregate throughput as follows
$$S(n) = \frac{p^{(n)}_S\cdot E[P]}{p^{(n)}_IT_I + p^{(n)}_ST_S + p^{(n)}_CT_C}$$
\begin{figure}[h]
\centering
\includegraphics[scale=0.25,angle=-90]{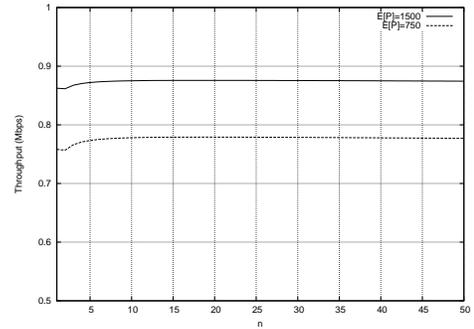}
\caption {Variation of throughput with the number of contending nodes. For the above plot, the data rate is set at $1\,Mbps$ and the others parameters were set to mimic 802.11b}
\label{fig:thruput}
\end{figure}
\subsection{Random polling system framework (RPS)}
From Figure \ref{fig:thruput}, it can be seen that the aggregate service rate is nearly constant. This suggests that we can imagine a constant-rate server serving the collection of queues. Accordingly, we consider the system as $n$ infinite-buffer queues being served by a single server. When the server visits a non empty queue, it serves one packet and moves on to the next queue. This type of polling system are is referred to as a \emph{1-Limited Random Polling System}. The service time of customers in queue $i$ is a nonnegative random variable with mean $p_i$ and second moment $p^{(2)}_i$. We define the expected offered load to the system due to queue $i$ as $\rho_i = a_i p_i$. The expected system utilization factor is defined as $\rho = \sum^{n}_{i=1} a_i p_i$. Clearly, a necessary condition for the system to reach a stable state is 
$$\rho < 1$$ 

Now we restate the notations and assumptions used in \cite{lee}, merely for the purpose of understanding our approach. A memoryless polling policy is assumed, such that the next queue, say queue $j$, is selected for service with probability $\gamma_j$, where $0 < \gamma_j < 1$. The service time provided (if any) and the switchover time that follows are collectively defined as a \emph{period}. The switchover time after the server visits queue $i$ is a nonnegative random variable with mean $s_i$ and second moment $s^{(2)}_i$. Let $b_i$ and $b^{(2)}_i$ be the first and second moment of the batch size. Let $s = \sum^{n}_{j=1} s_j \gamma_j$.

Let $Q_i(k)$ denote the queue length at node $i$, at the beginning of the $k^{th}$ \emph{period}. Then we define $E[Q_i] = \displaystyle \lim_{k \to \infty} E[Q_i(k)]$. From \textit{Theorem 12} of \cite{lee}, we can obtain a closed form expression for the average waiting time of the packets at node $i$ as
\begin{eqnarray} 
E[W_i] = \frac{E[Q_i]}{a_i \cdot P\{Q_i \geq 1 \}} - \left(\frac{1-\rho_i}{a_i}\right) -\frac{b^{(2)}_{i}- b_{i}}{2a_i b_{i}} \label{eq:ewi} \\
E[Q_i] = \frac{\psi_i}{\chi_i} + \frac{s  a^2_i}{2\gamma_i (1-\rho) \chi_i} \cdot \frac{\sum^{n}_{l=1} \displaystyle \frac{p_l \psi_l}{\chi_l}}{1 - \sum^{n}_{l=1} \displaystyle \frac{p_l a^2_l s}{2\gamma_l(1-\rho)\chi_l}}  \label{eq:eqi}
\end{eqnarray}
where
\begin{eqnarray} 
&& \chi_i = 1 - \frac{s a_i}{\gamma_i} - \frac{\rho s a_i}{2\gamma_i(1-\rho)} \label{eq:chi} \nonumber \\
&& \nabla_{ij} = a_i a_j \sum^{n}_{j=1} \left[ \gamma_j s^{(2)}_j + (p^{(2)}_j + 2s_jp_j)\frac{a_j s}{(1-\rho)} \right] \label{eq:nabla} \\
&& +\frac{s (e_{ij}+a_i\boldsymbol 1_{\{ i=j\}})}{1-\rho}- \frac{a_ia_js (s_i + s_j + p_i + p_j)}{(1-\rho)} \nonumber \\
&& \psi_i = \frac{\nabla_{ii}}{2 \gamma_i} + \frac{a_i}{2\gamma_i(1-\rho)} \cdot \sum^{n}_{l=1} p_l \nabla_{il} \label{eq:psi} \nonumber 
\end{eqnarray}
The steady state probability of a queue being non-empty is given by
\begin{eqnarray} \label{eq:pqi}
P\{Q_i \geq 1 \} = \frac{sa_i}{\gamma_i(1-\rho)} 
\end{eqnarray}
The above analysis applies only for a system with non-zero switch over times. 

\subsection{Application of RPS to a Single Cell 802.11 WLAN}
According to our modelling assumptions, if the server visits a node with a nonempty queue, it will serve exactly one packet and will move onto the next queue in zero switchover times. Else, if the server visits a node with an empty queue, we assume that it will move onto the next queue with zero switchover times. As 802.11 DCF follows fair server allocation policy, we can say that $\{ \gamma_i=\frac{1}{n} \, \forall \, 1\leq i\leq n\}$. This emulates the process of nonempty queues succeeding with equal probability. So we are interested in simplifying Equation \eqref{eq:ewi} to reflect the case of zero length switchover period. 

In a different context, the author in \cite{levy} has proposed that the expression for mean delay with zero switchover times can be obtained from the expression for non-zero switchover times by proper application of limits to the distribution of the switchover times. Motivated by this, we follow a similar approach to arrive at the expression for zero switchover times by defining the switchover times after servicing queue $i$ (i.e $s_i$) as a constant $\epsilon$. The expression for average delay when \emph{switch over times are zero} ($E[W^0_i]$) can be obtained by letting $\epsilon$ go to $0$. Thus we have 
\begin{eqnarray} \label{eq:ewi1}
E[W^0_i] =   \frac{\rho_i}{a_i}  - \frac{1}{a_i} + \lim_{\epsilon \to 0} \frac{E[Q_i]}{a_i \cdot P\{Q_i \geq 1 \}}-\frac{b^{(2)}_{i}-b_{i}}{2a_ib_{i}}
\end{eqnarray}
Since we have uncorrelated arrival processes, we have $\{e_{ij}=0, \forall i \neq j\}$. Now, after evaluating the limit in Equation \eqref{eq:ewi1} and through some algebraic manipulation, we obtain
\begin{eqnarray} \label{eq:delay}
E[W^0_i] &=&  \frac{\sum^n_{l=1} (p^{(2)}_l - p^2_l) a_l}{2(1-\rho)} + \frac{p_i}{2}\left(1 + \frac{e_{ii}}{a_i(1-\rho)} \right) \nonumber + \\
&& \frac{e_{ii}}{2a_i} \left( \frac{e_{ii}}{a_i} - \frac{b^{(2)}_i}{b_i}\right)
\end{eqnarray}

\subsection{Computation of $e_{ii}$ and $a_i$ for a Batch Poisson process}
The expression in \eqref{eq:delay} is valid only for class of processes that have the \emph{stationary and independent increment} property. Since the batch Poisson process is a member of the same class, we use batch Poisson process to model variable packet lengths. Therefore, in order to model the delay for a packet of variable length, we consider as a batch of packets that arrive at the node at the same time. 

We invoke the linear quadratic model of uncertainty from \cite{lee} to compute $e_{ii}$ and $a_i$. The motivation and the example for the same have been well described in the same paper. The model is as follows.
\begin{eqnarray} 
E[A_i(t)]  &=& a_i t \label{eq:lum1} \\
E[A_i(t)^2]  &=& e_{ii}t + a^2_it^2 \label{eq:lum2} 
\end{eqnarray}
where $A_i(t)$ is the number of packets that arrive at node $i$ during the time interval $[0,t]$ Since the inter-arrival times between the batches are exponentially distributed random variables and the batch size is independent of the inter-arrival times, hence $A_i(t)$ satisfies stationary independent increment property. Let $N_{i}(t)$ be the number of batches that have arrived during the time interval $[0, t]$ at node $i$. Let $B_{ij}$ be the random variable representing the batch size of the $j^{th}$ batch of this arrival process with mean $b_i$ and second moment $b^{(2)}_i$. Let the batch arrival process at node $i$ be Poisson with rate $\lambda_i$. Then, the moment generating function (MGF) for the batch Poisson Process is
$$M_{A_i(t)}(z) = E[z^{A_i(t)}] =  E[z^{\sum^{N_i(t)}_{j=1} B_{ij}}]$$
The random variables $\{B_{ij} , j \geq 0\}$ are i.i.d. Using the properties of moment generating function, we obtain
\begin{eqnarray}
a_i = \lambda_i M^{'}_{B_i(t)}(1)=\lambda_i b_i \\ \label{eq:ai}
e_{ii} = \lambda_i M^{''}_{B_i(t)}(1)=\lambda_i b^{(2)}_i \label{eq:eii} 
\end{eqnarray}

\section{\large{Application Delay Modelling}}
\label{analysis}
In this section we apply the framework laid in the previous section to arrive at closed form expression for the mean delay of an application packet. In our analysis, we allow the application packet length to vary from zero to any arbitrary length. But in practice, packets whose lengths are larger than the Maximum Transmission Unit (MTU), are split and sent as different entities. We are interested in the overall delay of a packet, which may compose of different smaller MTUs. Thus we quantify the mean packet delay, by analyzing two separate cases.

\subsection{Application packet size is bounded above by the MTU}
Let us assume that the MTU has a length of $P\, bits$. All other packet lengths are expressed in terms of the MTU. At node $i$, we define random variable $\Omega_i=\frac{P_i}{P}$ with mean $\omega_i$ and second moment $\omega^{(2)}_i$. We would like to emphasis the fact that the random variable $\Omega_i$ will be a fraction. 

In order to obtain the expression for delay when the packet size is bounded above by the size of the \textit{MTU}, we set 
$$ p_i = \frac{\omega_iP}{C}	\quad \textrm{and} \quad p^{(2)}_i = \frac{\omega^{(2)}_iP^2}{C^2} $$ 
Since the packet size equals a MTU at most, we also set the batch size to a a fixed value 1, thus we get
$$ b_i = 1	\quad \textrm{and} \quad b^{(2)}_i = 1 $$  
Setting the above variables in Equation \eqref{eq:delay}, we get the expression for average delay as
\begin{eqnarray} \label{eq:delay1}
d_{avg} =  \frac{\sum^n_{l=1} (\omega^{(2)}_l - \omega^2_l) \lambda_l}{2 \left( \frac{C}{P} \right) ^2(1-\rho)} + \frac{\omega_i}{2\left( \frac{C}{P} \right)}\left(\frac{2-\rho}{1-\rho} \right)
\end{eqnarray}

\subsection{Application packet size is bounded below by the MTU}
As in the previous section, we assume the unit packet size to be $P\, bits$ and that every other packet length is expressed in terms of this unit packet. In this section, we take an alternate approach as compared to the one in the previous section. We model the large packet as an aggregation of smaller units of packet, each of which has size lesser than or equal to $P\, bits$. For a node $i$, we model this aggregated packet as a batch Poisson process, whose batch size has mean $\omega_i$ and second moment $\omega^{(2)}_i$; we do not restrict the batch size to be discrete. As the packet size is greater than \textit{MTU}, the packet is fragmented at the source and reassembled at the destination. We note that the packet is constructed only after the reception of its fragments. Or on other words, we can say that the mean packet delay is equal to the mean delay of the last fragment. In this section, we are interested in quantifying the total delay of all the fragments of a packet. 

Since the packet size is greater than the \textit{MTU}, the packet is fragmented at the source.  Thus, this effect had to be factored into the computation of aggregate throughput ($C\, bits/s$). Due to lack of space, we are not presenting the modified expression in this paper. From the arguments in the section \ref{bia}, we assume  that an MTU at node $i$ is serviced at a constant time of $\frac{P}{C}$ sec\footnotemark \footnotetext{We assume that fragments having size lesser than MTU are also transmitted at same rate, thus our expression is in-fact an upper bound for the mean delay experienced by a packet. Simulations validate our assumption by showing that the bound is in-fact tight.}. By adopting the throughput computation by \cite{bianchi}, we have abstracted the idle and the collision slots into the constant service time for a packet of node $i$, given by $\frac{P}{C}$ sec. We model our system using the RPS framework by taking a server which serves at fixed rate. Thus,
$$\forall i,\, p_i = \frac{P}{C}  	 \textrm{ and } p^{(2)}_i = \left(\frac{P}{C}\right)^2$$
By substituting the various values in \eqref{eq:delay}, as in the previous section, we get the mean delay for a fragment of the packet as 
\begin{eqnarray} \label{eq:delay1}
d_i =   \frac{P}{2C}\left(1 + \frac{\omega^{(2)}_i}{\omega_i(1-\rho)} \right)
\end{eqnarray}
This fragment is held at the network buffer at the destination node, until the remaining fragment arrive. Now the we approximate the average cycle time of the server at node $i$ as
$$  w_i = d_i - \frac{P}{C}$$
It can be shown that the mean packet delay is related to the mean fragment delay and the mean batch size as follows,
$$d_{avg} = \left( \frac{\omega_i+1}{2} \right)d_i+ \left(\frac{\omega_i-1}{2} \right)\frac{P}{C}$$
After substituting the value of $d_i$, we obtain the average delay per packet as
\begin{eqnarray} \label{eq:delay2}
d_{avg} = \frac{P}{4C} \cdot \left(3 - \omega_i + \frac{\omega^{(2)}_i(1+\omega_i)}{\omega_i(1-\rho)} \right)
\end{eqnarray}

\section{\large{Validation}}
\label{sim}
In this section, we show tables which compare the analytical against the simulated values of mean delay. The simulation was done using Qualnet 4.5, which is a discrete event simulation system. In order to obtain accurate estimates of the mean delay, the simulation was run long enough so that the average delay at the destinations are within $1 \mu s$ interval. This process was repeated for $30$ simulation. In the tables, one packet is equivalent to $1500\, Bytes$ (i.e MTU = $1500\, Bytes$ ). Data Rate is $1 \, Mbps$ and  $4$ nodes contend for the WLAN medium.

\begin{table}[!h]
\caption{Table of delay for simulation and the analytical values for $\rho \approx \, 0.42$. Packet size is distributed uniformly in $[750, 1500) \, Bytes$. $C = 70.0\, pkts/sec$ }
\begin{center}
\begin{tabular}{|c|c|c|c|c|c|c|c|c|}
\hline $\lambda_1 $ & $\lambda_2$ & $\lambda_3$ & $\lambda_4$ & $d^{(1)}_{avg}$ & $d^{(2)}_{avg}$ & $d^{(3)}_{avg}$ & $d^{(4)}_{avg}$ & $d_{avg}$  \\ 
\hline 10.0 & 10.0 & 10.0 & 10.0 & 15.0 & 15.0 & 14.9 & 14.9 & 14.9 \\ 
\hline 2.0 & 12.7 & 12.7 & 12.7 & 14.7 & 14.6 & 14.7 & 14.7 & 14.9 \\ 
\hline 2.0 & 2.0 & 2.0 & 34.5 & 14.4 & 14.5 & 14.4 & 14.8 &  15.0\\ 
\hline
\end{tabular} 
\end{center}
\label{table1}
\end{table}

\begin{table}[!h]
\caption{Table of delay for simulation and the analytical values for $\rho \approx \, 0.71$. Packet size is distributed uniformly in $[750, 1500) \, Bytes$. $C = 70.0\, pkts/sec$ }
\begin{center}
\begin{tabular}{|c|c|c|c|c|c|c|c|c|}
\hline $\lambda_1 $ & $\lambda_2$ & $\lambda_3$ & $\lambda_4$ & $d^{(1)}_{avg}$ & $d^{(2)}_{avg}$ & $d^{(3)}_{avg}$ & $d^{(4)}_{avg}$ & $d_{avg}$  \\ 
\hline 16.7 & 16.7 & 16.7 & 16.7 & 24.7 & 24.7 & 24.6 & 24.7 & 24.6 \\ 
\hline  2.0 & 21.3 & 21.3 & 21.3 & 21.0 & 23.4 & 23.6 & 23.3 & 24.0 \\ 
\hline  2.0 & 2.0 & 2.0 & 58.8 & 18.5 & 18.5 & 18.6 & 23.6 & 23.3 \\ 
\hline 
\end{tabular} 
\end{center}
\label{table2}
\end{table}

In Tables \ref{table1} , \ref{table2}, \ref{table3} and \ref{table4}, we compares the analytical and simulation result for various arrival rate and type of distribution for the packet size for the case when the packets size is less than the \textit{MTU}. We note that for lower aggregate rate, the mean delay across the queues are invariant and matches with the analytical results. For higher load, our analysis is able to approximate the maximum delay experienced by a packet in the network.

\begin{table}[!h]
\caption{Table of delay for simulation and the analytical values for $\rho \approx \, 0.28$. Packet size is distributed exponentially with mean $1125 \, Bytes$. $C = 69.2\, pkts/sec$ }
\begin{center}
\begin{tabular}{|c|c|c|c|c|c|c|c|c|}
\hline $\lambda_1 $ & $\lambda_2$ & $\lambda_3$ & $\lambda_4$ & $d^{(1)}_{avg}$ & $d^{(2)}_{avg}$ & $d^{(3)}_{avg}$ & $d^{(4)}_{avg}$ & $d_{avg}$  \\ 
\hline  6.7 & 6.7 & 6.7 & 6.7 & 15.7 & 15.7 & 15.9 & 15.8 & 15.1 \\ 
\hline  2.0 & 8.3 & 8.3 & 8.3 & 15.7 & 15.7 & 15.7 & 15.7 & 15.2 \\  
\hline  2.0 & 2.0 & 2.0 & 20.8 & 15.3 & 15.1 & 15.2 & 15.7 & 15.2 \\ 
\hline 
\end{tabular} 
\end{center}
\label{table3}
\end{table}

\begin{table}[!h]
\caption{Table of delay for simulation and the analytical values for $\rho \approx \, 0.57$. Packet size is distributed exponentially with mean $1125 \, Bytes$. $C = 69.2\, pkts/sec$ }
\begin{center}
\begin{tabular}{|c|c|c|c|c|c|c|c|c|}
\hline $\lambda_1 $ & $\lambda_2$ & $\lambda_3$ & $\lambda_4$ & $d^{(1)}_{avg}$ & $d^{(2)}_{avg}$ & $d^{(3)}_{avg}$ & $d^{(4)}_{avg}$ & $d_{avg}$  \\ 
\hline  13.3 & 13.3 & 13.3 & 13.3 & 26.9 & 26.9 & 27.0 & 27.6 & 25.6 \\  
\hline  2.0 & 17.2 & 17.2 & 17.2 & 23.3 & 26.8 & 27.2 & 27.3 & 25.8 \\ 
\hline  2.0 &  2.0 & 2.0  & 49.6 & 20.3 & 20.0 & 20.3 & 28.2 & 25.7 \\ 
\hline 
\end{tabular} 
\end{center}
\label{table4}
\end{table}

\begin{table}[!h]
\caption{Table of delay for simulation and the analytical values for $\rho \approx \, 0.19$. Packet size is distributed uniformly in $[1500, 4500) \, Bytes$. $C = 68.9\, pkts/sec$ }
\begin{center}
\begin{tabular}{|c|c|c|c|c|c|c|c|c|}
\hline $\lambda_1 $ & $\lambda_2$ & $\lambda_3$ & $\lambda_4$ & $d^{(1)}_{avg}$ & $d^{(2)}_{avg}$ & $d^{(3)}_{avg}$ & $d^{(4)}_{avg}$ & $d_{avg}$  \\ 
\hline  1.7 & 1.7 & 1.7 & 1.7 & 35.3 & 34.8 & 34.9 & 35.2 & 32.9 \\ 
\hline  1.0 & 1.9 & 1.9 & 1.9 & 33.9 & 32.8 & 33.0 & 32.9 & 32.9 \\ 
\hline  1.0 & 1.0 & 1.0 & 3.7 & 34.5 & 34.1 & 34.0 & 32.5 & 32.4 \\ 
\hline 
\end{tabular} 
\end{center}
\label{table5}
\end{table}

\begin{table}[!h]
\caption{Table of delay for simulation and the analytical values for $\rho \approx \, 0.58$. Packet size is distributed uniformly in $[1500, 4500) \, Bytes$. $C = 69.8\, pkts/sec$ }
\begin{center}
\begin{tabular}{|c|c|c|c|c|c|c|c|c|}
\hline $\lambda_1 $ & $\lambda_2$ & $\lambda_3$ & $\lambda_4$ & $d^{(1)}_{avg}$ & $d^{(2)}_{avg}$ & $d^{(3)}_{avg}$ & $d^{(4)}_{avg}$ & $d_{avg}$  \\ 
\hline  5.0 & 5.0 & 5.0 & 5.0 & 52.9 & 52.7 & 52.2 & 52.3 & 59.8 \\ 
\hline  1.0 & 6.4 & 6.4 & 6.4 & 52.3 & 52.3 & 52.5 & 52.9 & 60.0 \\ 
\hline  1.0 & 1.0 & 1.0 & 16.9 & 47.7 & 47.6 & 47.4 & 50.0 & 59.6 \\ 
\hline 
\end{tabular} 
\end{center}
\label{table6}
\end{table}

\begin{table}[!h]
\caption{Table of delay for simulation and the analytical values for $\rho \approx \, 0.19$. Packet size is distributed exponentially with mean $3000 \, Bytes$. $C = 62.5\, pkts/sec$ }
\begin{center}
\begin{tabular}{|c|c|c|c|c|c|c|c|c|}
\hline $\lambda_1 $ & $\lambda_2$ & $\lambda_3$ & $\lambda_4$ & $d^{(1)}_{avg}$ & $d^{(2)}_{avg}$ & $d^{(3)}_{avg}$ & $d^{(4)}_{avg}$ & $d_{avg}$  \\ 
\hline  1.7 & 1.7 & 1.7 & 1.7 & 35.3 & 34.8 & 34.9 & 35.1 & 34.5 \\ 
\hline  1.0 & 1.9 & 1.9 & 1.9 & 35.3 & 34.7 & 34.2 & 36.0 & 34.5 \\ 
\hline  1.0 & 1.0 & 1.0 & 3.7 & 35.1 & 35.9 & 36.1 & 34.9 & 34.5 \\ 
\hline 
\end{tabular} 
\end{center}
\label{table7}
\end{table}

\begin{table}[!h]
\caption{Table of delay for simulation and the analytical values for $\rho \approx \, 0.62$. Packet size is distributed exponentially with mean $3000 \, Bytes$. $C = 62.5\, pkts/sec$ }
\begin{center}
\begin{tabular}{|c|c|c|c|c|c|c|c|c|}
\hline $\lambda_1 $ & $\lambda_2$ & $\lambda_3$ & $\lambda_4$ & $d^{(1)}_{avg}$ & $d^{(2)}_{avg}$ & $d^{(3)}_{avg}$ & $d^{(4)}_{avg}$ & $d_{avg}$  \\ 
\hline  5.0 & 5.0 & 5.0 & 5.0 & 68.3 & 64.2 & 63.2 & 65.0 & 70.7 \\ 
\hline  1.0 & 6.4 & 6.4 & 6.4 & 56.1 & 63.7 & 64.2 & 64.1 & 71.3 \\ 
\hline  1.0 & 1.0 & 1.0 & 16.9 & 50.1 & 50.7 & 51.0 & 66.5 & 70.1 \\ 
\hline 
\end{tabular} 
\end{center}
\label{table8}
\end{table}

In Tables \ref{table5} , \ref{table6}, \ref{table7} and \ref{table8}, we compare the analytical and simulation result for various arrival rate and type of distribution for the packet size for the case when the packets size is greater that the \textit{MTU}. We note that, for the case of homogenous arrival, the analysis is matching with the simulation result  under high loads as well.

\section{\large{Conclusion}}
In this paper, we have the abstracted the mechanism of IEEE 802.11 MAC enabling us to find simple closed form expression for application level packet delay in a single cell IEEE 802.11 wireless area network. Our analysis enables to approximate the application level mean delay for variable packet length using a closed form expressions. For low loads, the delay across the queues is invariant and is closely approximated by the analytical formula. It can be seen that, even for the scenario of nonhomogeneous packet arrivals, the analytical delays is closest to the mean delay of a packet in the queue with the highest data rate. Simulations indicate that the proposed framework is able to model the maximum mean delay in single hop wireless mesh network with reasonable accuracy, provided the system operates within the capacity region.

\bibliography{delay_modelling_rpsicc_sep27}

\begin{thebibliography}{10}
\providecommand{\url}[1]{#1}
\csname url@samestyle\endcsname
\providecommand{\newblock}{\relax}
\providecommand{\bibinfo}[2]{#2}
\providecommand{\BIBentrySTDinterwordspacing}{\spaceskip=0pt\relax}
\providecommand{\BIBentryALTinterwordstretchfactor}{4}
\providecommand{\BIBentryALTinterwordspacing}{\spaceskip=\fontdimen2\font plus
\BIBentryALTinterwordstretchfactor\fontdimen3\font minus
  \fontdimen4\font\relax}
\providecommand{\BIBforeignlanguage}[2]{{%
\expandafter\ifx\csname l@#1\endcsname\relax
\typeout{** WARNING: IEEEtran.bst: No hyphenation pattern has been}%
\typeout{** loaded for the language `#1'. Using the pattern for}%
\typeout{** the default language instead.}%
\else
\language=\csname l@#1\endcsname
\fi
#2}}
\providecommand{\BIBdecl}{\relax}
\BIBdecl

\bibitem{bianchi}
{G. Bianchi}, ``Performance analysis of the \textsc{IEEE 802.11} distributed
  coordination function,'' in \emph{JSAC Wireless Series, vol. 18, no. 3},
  2000.

\bibitem{tay}
{Y. Tay} and {K. Chua}, ``A capacity analysis for the \textsc{IEEE 802.11 MAC}
  protocol,'' in \emph{Wireless Networks, vol. 7, no. 2, pp. 159-171}, March
  2001.

\bibitem{akumar}
{A. Kumar}, {E. Altman}, {D. Miorandi}, and {M. Goyal}, ``New insights from a
  fixed point analysis of single cell \textsc{IEEE 802.11 WLAN}s,'' in
  \emph{Proceedings of IEEE INFOCOM}, 2005.

\bibitem{tobagi}
{K. Medepalli} and {F.A. Tobagi}, ``System centric and user centric queueing
  models for \textsc{IEEE 802.11} based wireless \textsc{LAN}s,'' in
  \emph{Proc. of IEEE 2nd International Conference on Broadband Networks}, San
  Francisco, October 2005.

\bibitem{panda}
{Manoj K. Panda} and {Anurag Kumar}, ``State dependent attempt rate modeling of
  single cell \textsc{IEEE} 802.11 \textsc{WLAN}s with homogeneous nodes and
  poisson arrivals,'' in \emph{Proceedings of IEEE COMSNET}, 2009.

\bibitem{joy}
{Albert Sunny}, {Joy Kuri}, and {Saurabh Aggarwal}, ``Delay modelling for a
  single-hop wireless mesh network under light aggregate traffic,'' online
  version available at \emph{http://www.arxiv.org/abs/1009.0448}.

\bibitem{tickoo}
{O. Tickoo} and {B. Sikdar}, ``Queueing analysis and delay mitigation in
  \textsc{IEEE 802.11} random access mac based wireless networks,'' in
  \emph{Proc. of IEEE INFOCOM}, 2004.

\bibitem{saurabh}
{Albert Sunny}, {Joy Kuri}, and {Saurabh Aggarwal}, ``Delay modelling for
  single cell \textsc{IEEE 802.11} wlans using a random polling system,''
  online version available at \emph{http://www.arxiv.org/abs/1009.3468}.

\bibitem{lee}
{Thomas (Yew Sing) Lee}, ``A closed form solution for the asymmetric random
  polling system with correlated levy input process,'' in \emph{Mathematics of
  Operations Research, Vol. 22, No. 2 (May, 1997), pp. 432-457}.

\bibitem{levy}
{Leonard Kleinrock} and {Hanoch Levy}, ``The analysis of random polling
  systems,'' in \emph{Operations Research, Vol. 36, No. 5 (Sep - Oct) , pp.
  716-732}, 1988.

\end{thebibliography}

\end{document}